\pdfoutput=1
\documentclass[aps,prd,twocolumn,superscriptaddress,preprintnumbers,nofootinbib]{revtex4}
\usepackage{amsmath}
\usepackage{amssymb}
\usepackage{natbib}
\usepackage{graphicx}
\usepackage{epsf}
\usepackage{subfigure}
\usepackage{color}
\usepackage{threeparttable}
\usepackage{comment}
\usepackage{epsfig}
\usepackage{xspace}
\usepackage{hyperref}
\usepackage{latexsym}
\graphicspath{{figures/}{./}}
\DeclareGraphicsExtensions{.jpg,.pdf,.png,.eps,.ps}

 \newcommand{\lcdm}{\mbox{$\Lambda$CDM}\xspace}

 \newcommand{\ltsima}{$\; \buildrel < \over \sim \;$}
 \newcommand{\ltsim}{\lower.5ex\hbox{\ltsima}}

 \newcommand{\sqdeg}{\ensuremath{\mathrm{deg}^2}}

 \newcommand{\bmode}{\ensuremath{B}-mode}
 \newcommand{\bmodes}{\ensuremath{B}-modes}
  \newcommand{\emode}{\ensuremath{E}-mode}
 \newcommand{\emodes}{\ensuremath{E}-modes}
 \newcommand{\alens}{\ensuremath{A_\mathrm{lens}}}

 \newcommand{\muksq}{\ensuremath{\mu{\rm K}^2}}
 
 \newcommand{\wmap}{\textit{WMAP}\xspace}
 
 \newcommand{\planck}{\textit{Planck}}
 \newcommand{\sptpol}{{SPTpol}}

 \newcommand{\simleq}{{\raise.0ex\hbox{$\mathchar"013C$}\mkern-14mu \lower1.2ex\hbox{$\mathchar"0218$}}}
 \newcommand{\simgeq}{{\raise.0ex\hbox{$\mathchar"013E$}\mkern-14mu \lower1.2ex\hbox{$\mathchar"0218$}}}
\newcommand{\be}{\begin{equation}}
\newcommand{\ee}{\end{equation}}
\newcommand{\bea}{\begin{eqnarray}}
\newcommand{\eea}{\end{eqnarray}}

\newcommand{\EtoB}{$E \rightarrow B$}

\newcommand{\Comment}[1]{}
\renewcommand{\deg}{\ensuremath{^\circ}}

\newcommand{\abbrzero}{\ensuremath{ 1.17 \pm  0.13}}
\newcommand{\abbhifg}{\ensuremath{ 1.14 \pm  0.13}}
\newcommand{\abblofg}{\ensuremath{ 1.18 \pm  0.13}}
\newcommand{\abbnoninety}{\ensuremath{ 1.13 \pm  0.13}}

\newcommand{\sigmaabbnoonefifty}{\ensuremath{ 0.25}}
\newcommand{\rlimit}{\ensuremath{ 0.44}}
\newcommand{\rlimitnoninety}{\ensuremath{ 0.44}}
\newcommand{\rlimitnoonefifty}{\ensuremath{ 1.40}}
\newcommand{\rlimithifg}{\ensuremath{ 0.39}}
\newcommand{\rlimitlofg}{\ensuremath{ 0.46}}
\newcommand{\rlimitnogalprior}{\ensuremath{ 0.43}}
\newcommand{\apmf}{\ensuremath{ 0.37}}
\newcommand{\apmflobeta}{\ensuremath{ 0.42}}

\newcommand{\positiveBBpte}{\ensuremath{5.8\times 10^{-71}}}
\newcommand{\positiveBBsignif}{18.1} 

\newcommand{\positiveBBwSVFGpte}{\ensuremath{1.8\times 10^{-18}}}
\newcommand{\positiveBBwSVFGsignif}{8.7} 
\newcommand{\positiveBBwFGpte}{\ensuremath{2.6\times 10^{-29}}}
\newcommand{\positiveBBwFGsignif}{11.2} 

\begin{document}
\title{Measurements of \bmode{} Polarization of the Cosmic Microwave Background from 500 Square Degrees of SPTpol Data}

\affiliation{Department of Astrophysical and Planetary Sciences, University of Colorado, Boulder, CO, USA 80309}
\affiliation{Department of Physics, University of Colorado, Boulder, CO, USA 80309}
\affiliation{Physics Department, Center for Education and Research in Cosmology and Astrophysics, Case Western Reserve University, Cleveland, OH, USA 44106}
\affiliation{School of Physics, University of Melbourne, Parkville, VIC 3010, Australia}
\affiliation{High Energy Physics Division, Argonne National Laboratory, 9700 S. Cass Avenue, Argonne, IL, USA 60439}
\affiliation{Kavli Institute for Cosmological Physics, University of Chicago, 5640 South Ellis Avenue, Chicago, IL, USA 60637}
\affiliation{Cardiff University, Cardiff CF10 3XQ, United Kingdom}
\affiliation{Fermi National Accelerator Laboratory, MS209, P.O. Box 500, Batavia, IL 60510}
\affiliation{NIST Quantum Devices Group, 325 Broadway Mailcode 817.03, Boulder, CO, USA 80305}
\affiliation{Department of Physics, University of California, Berkeley, CA, USA 94720}
\affiliation{Department of Astronomy and Astrophysics, University of Chicago, 5640 South Ellis Avenue, Chicago, IL, USA 60637}
\affiliation{Department of Physics, University of Chicago, 5640 South Ellis Avenue, Chicago, IL, USA 60637}
\affiliation{Enrico Fermi Institute, University of Chicago, 5640 South Ellis Avenue, Chicago, IL, USA 60637}
\affiliation{Department of Physics, McGill University, 3600 Rue University, Montreal, Quebec H3A 2T8, Canada}
\affiliation{School of Mathematics, Statistics \& Computer Science, University of KwaZulu-Natal, Durban, South Africa}
\affiliation{University of Chicago, 5640 South Ellis Avenue, Chicago, IL, USA 60637}
\affiliation{TAPIR, Walter Burke Institute for Theoretical Physics, California Institute of Technology, 1200 E California Blvd, Pasadena, CA, USA 91125}
\affiliation{California Institute of Technology, MS 249-17, 1216 E. California Blvd., Pasadena, CA, USA 91125}
\affiliation{Physics Division, Lawrence Berkeley National Laboratory, Berkeley, CA, USA 94720}
\affiliation{Canadian Institute for Advanced Research, CIFAR Program in Gravity and the Extreme Universe, Toronto, ON, M5G 1Z8, Canada}
\affiliation{Harvey Mudd College, 301 Platt Blvd., Claremont, CA 91711}
\affiliation{European Southern Observatory, Karl-Schwarzschild-Str. 2, 85748 Garching bei M\"{u}nchen, Germany}
\affiliation{Astronomy Department, University of Illinois at Urbana-Champaign, 1002 W. Green Street, Urbana, IL 61801, USA}
\affiliation{Department of Physics, University of Illinois Urbana-Champaign, 1110 W. Green Street, Urbana, IL 61801, USA}
\affiliation{SLAC National Accelerator Laboratory, 2575 Sand Hill Road, Menlo Park, CA 94025}
\affiliation{Dept. of Physics, Stanford University, 382 Via Pueblo Mall, Stanford, CA 94305}
\affiliation{Department of Physics, University of California, One Shields Avenue, Davis, CA, USA 95616}
\affiliation{Department of Physics, University of Michigan, 450 Church Street, Ann  Arbor, MI, USA 48109}
\affiliation{Dunlap Institute for Astronomy \& Astrophysics, University of Toronto, 50 St George St, Toronto, ON, M5S 3H4, Canada}
\affiliation{Materials Sciences Division, Argonne National Laboratory, 9700 S. Cass Avenue, Argonne, IL, USA 60439}
\affiliation{School of Physics and Astronomy, University of Minnesota, 116 Church Street S.E. Minneapolis, MN, USA 55455}
\affiliation{Department of Physics, Yale University, P.O. Box 208120, New Haven, CT 06520-8120}
\affiliation{Liberal Arts Department, School of the Art Institute of Chicago, 112 S Michigan Ave, Chicago, IL, USA 60603}
\affiliation{Three-Speed Logic, Inc., Victoria, B.C., V8S 3Z5, Canada}
\affiliation{Harvard-Smithsonian Center for Astrophysics, 60 Garden Street, Cambridge, MA, USA 02138}
\affiliation{Department of Astronomy \& Astrophysics, University of Toronto, 50 St George St, Toronto, ON, M5S 3H4, Canada}
\affiliation{Department of Astronomy, University of Maryland College Park, MD, USA 20742}
\affiliation{Department of Physics and Astronomy, University of California, Los Angeles, CA, USA 90095}
\author{J.T.~Sayre}
\affiliation{Department of Astrophysical and Planetary Sciences, University of Colorado, Boulder, CO, USA 80309}
\affiliation{Department of Physics, University of Colorado, Boulder, CO, USA 80309}
\affiliation{Physics Department, Center for Education and Research in Cosmology and Astrophysics, Case Western Reserve University, Cleveland, OH, USA 44106}
\author{C.~L.~Reichardt}
\affiliation{School of Physics, University of Melbourne, Parkville, VIC 3010, Australia}
\author{J.~W.~Henning}
\affiliation{High Energy Physics Division, Argonne National Laboratory, 9700 S. Cass Avenue, Argonne, IL, USA 60439}
\affiliation{Kavli Institute for Cosmological Physics, University of Chicago, 5640 South Ellis Avenue, Chicago, IL, USA 60637}
\author{P.~A.~R.~Ade}
\affiliation{Cardiff University, Cardiff CF10 3XQ, United Kingdom}
\author{A.~J.~Anderson}
\affiliation{Fermi National Accelerator Laboratory, MS209, P.O. Box 500, Batavia, IL 60510}
\author{J.~E.~Austermann}
\affiliation{NIST Quantum Devices Group, 325 Broadway Mailcode 817.03, Boulder, CO, USA 80305}
\affiliation{Department of Physics, University of Colorado, Boulder, CO, USA 80309}
\author{J.~S.~Avva}
\affiliation{Department of Physics, University of California, Berkeley, CA, USA 94720}
\author{J.~A.~Beall}
\affiliation{NIST Quantum Devices Group, 325 Broadway Mailcode 817.03, Boulder, CO, USA 80305}
\author{A.~N.~Bender}
\affiliation{High Energy Physics Division, Argonne National Laboratory, 9700 S. Cass Avenue, Argonne, IL, USA 60439}
\affiliation{Kavli Institute for Cosmological Physics, University of Chicago, 5640 South Ellis Avenue, Chicago, IL, USA 60637}
\author{B.~A.~Benson}
\affiliation{Fermi National Accelerator Laboratory, MS209, P.O. Box 500, Batavia, IL 60510}
\affiliation{Kavli Institute for Cosmological Physics, University of Chicago, 5640 South Ellis Avenue, Chicago, IL, USA 60637}
\affiliation{Department of Astronomy and Astrophysics, University of Chicago, 5640 South Ellis Avenue, Chicago, IL, USA 60637}
\author{F.~Bianchini}
\affiliation{School of Physics, University of Melbourne, Parkville, VIC 3010, Australia}
\author{L.~E.~Bleem}
\affiliation{High Energy Physics Division, Argonne National Laboratory, 9700 S. Cass Avenue, Argonne, IL, USA 60439}
\affiliation{Kavli Institute for Cosmological Physics, University of Chicago, 5640 South Ellis Avenue, Chicago, IL, USA 60637}
\author{J.~E.~Carlstrom}
\affiliation{Kavli Institute for Cosmological Physics, University of Chicago, 5640 South Ellis Avenue, Chicago, IL, USA 60637}
\affiliation{Department of Physics, University of Chicago, 5640 South Ellis Avenue, Chicago, IL, USA 60637}
\affiliation{High Energy Physics Division, Argonne National Laboratory, 9700 S. Cass Avenue, Argonne, IL, USA 60439}
\affiliation{Department of Astronomy and Astrophysics, University of Chicago, 5640 South Ellis Avenue, Chicago, IL, USA 60637}
\affiliation{Enrico Fermi Institute, University of Chicago, 5640 South Ellis Avenue, Chicago, IL, USA 60637}
\author{C.~L.~Chang}
\affiliation{Kavli Institute for Cosmological Physics, University of Chicago, 5640 South Ellis Avenue, Chicago, IL, USA 60637}
\affiliation{High Energy Physics Division, Argonne National Laboratory, 9700 S. Cass Avenue, Argonne, IL, USA 60439}
\affiliation{Department of Astronomy and Astrophysics, University of Chicago, 5640 South Ellis Avenue, Chicago, IL, USA 60637}
\author{P.~Chaubal}
\affiliation{School of Physics, University of Melbourne, Parkville, VIC 3010, Australia}
\author{H.~C.~Chiang}
\affiliation{Department of Physics, McGill University, 3600 Rue University, Montreal, Quebec H3A 2T8, Canada}
\affiliation{School of Mathematics, Statistics \& Computer Science, University of KwaZulu-Natal, Durban, South Africa}
\author{R.~Citron}
\affiliation{University of Chicago, 5640 South Ellis Avenue, Chicago, IL, USA 60637}
\author{C.~Corbett~Moran}
\affiliation{TAPIR, Walter Burke Institute for Theoretical Physics, California Institute of Technology, 1200 E California Blvd, Pasadena, CA, USA 91125}
\author{T.~M.~Crawford}
\affiliation{Kavli Institute for Cosmological Physics, University of Chicago, 5640 South Ellis Avenue, Chicago, IL, USA 60637}
\affiliation{Department of Astronomy and Astrophysics, University of Chicago, 5640 South Ellis Avenue, Chicago, IL, USA 60637}
\author{A.~T.~Crites}
\affiliation{Kavli Institute for Cosmological Physics, University of Chicago, 5640 South Ellis Avenue, Chicago, IL, USA 60637}
\affiliation{Department of Astronomy and Astrophysics, University of Chicago, 5640 South Ellis Avenue, Chicago, IL, USA 60637}
\affiliation{California Institute of Technology, MS 249-17, 1216 E. California Blvd., Pasadena, CA, USA 91125}
\author{T.~de~Haan}
\affiliation{Department of Physics, University of California, Berkeley, CA, USA 94720}
\affiliation{Physics Division, Lawrence Berkeley National Laboratory, Berkeley, CA, USA 94720}
\author{M.~A.~Dobbs}
\affiliation{Department of Physics, McGill University, 3600 Rue University, Montreal, Quebec H3A 2T8, Canada}
\affiliation{Canadian Institute for Advanced Research, CIFAR Program in Gravity and the Extreme Universe, Toronto, ON, M5G 1Z8, Canada}
\author{W.~Everett}
\affiliation{Department of Astrophysical and Planetary Sciences, University of Colorado, Boulder, CO, USA 80309}
\author{J.~Gallicchio}
\affiliation{Kavli Institute for Cosmological Physics, University of Chicago, 5640 South Ellis Avenue, Chicago, IL, USA 60637}
\affiliation{Harvey Mudd College, 301 Platt Blvd., Claremont, CA 91711}
\author{E.~M.~George}
\affiliation{European Southern Observatory, Karl-Schwarzschild-Str. 2, 85748 Garching bei M\"{u}nchen, Germany}
\affiliation{Department of Physics, University of California, Berkeley, CA, USA 94720}
\author{A.~Gilbert}
\affiliation{Department of Physics, McGill University, 3600 Rue University, Montreal, Quebec H3A 2T8, Canada}
\author{N.~Gupta}
\affiliation{School of Physics, University of Melbourne, Parkville, VIC 3010, Australia}
\author{N.~W.~Halverson}
\affiliation{Department of Astrophysical and Planetary Sciences, University of Colorado, Boulder, CO, USA 80309}
\affiliation{Department of Physics, University of Colorado, Boulder, CO, USA 80309}
\author{N.~Harrington}
\affiliation{Department of Physics, University of California, Berkeley, CA, USA 94720}
\author{G.~C.~Hilton}
\affiliation{NIST Quantum Devices Group, 325 Broadway Mailcode 817.03, Boulder, CO, USA 80305}
\author{G.~P.~Holder}
\affiliation{Astronomy Department, University of Illinois at Urbana-Champaign, 1002 W. Green Street, Urbana, IL 61801, USA}
\affiliation{Department of Physics, University of Illinois Urbana-Champaign, 1110 W. Green Street, Urbana, IL 61801, USA}
\affiliation{Canadian Institute for Advanced Research, CIFAR Program in Gravity and the Extreme Universe, Toronto, ON, M5G 1Z8, Canada}
\author{W.~L.~Holzapfel}
\affiliation{Department of Physics, University of California, Berkeley, CA, USA 94720}
\author{J.~D.~Hrubes}
\affiliation{University of Chicago, 5640 South Ellis Avenue, Chicago, IL, USA 60637}
\author{N.~Huang}
\affiliation{Department of Physics, University of California, Berkeley, CA, USA 94720}
\author{J.~Hubmayr}
\affiliation{NIST Quantum Devices Group, 325 Broadway Mailcode 817.03, Boulder, CO, USA 80305}
\author{K.~D.~Irwin}
\affiliation{SLAC National Accelerator Laboratory, 2575 Sand Hill Road, Menlo Park, CA 94025}
\affiliation{Dept. of Physics, Stanford University, 382 Via Pueblo Mall, Stanford, CA 94305}
\author{L.~Knox}
\affiliation{Department of Physics, University of California, One Shields Avenue, Davis, CA, USA 95616}
\author{A.~T.~Lee}
\affiliation{Department of Physics, University of California, Berkeley, CA, USA 94720}
\affiliation{Physics Division, Lawrence Berkeley National Laboratory, Berkeley, CA, USA 94720}
\author{D.~Li}
\affiliation{NIST Quantum Devices Group, 325 Broadway Mailcode 817.03, Boulder, CO, USA 80305}
\affiliation{SLAC National Accelerator Laboratory, 2575 Sand Hill Road, Menlo Park, CA 94025}
\author{A.~Lowitz}
\affiliation{Department of Astronomy and Astrophysics, University of Chicago, 5640 South Ellis Avenue, Chicago, IL, USA 60637}
\author{J.~J.~McMahon}
\affiliation{Department of Physics, University of Michigan, 450 Church Street, Ann  Arbor, MI, USA 48109}
\author{S.~S.~Meyer}
\affiliation{Kavli Institute for Cosmological Physics, University of Chicago, 5640 South Ellis Avenue, Chicago, IL, USA 60637}
\affiliation{Department of Physics, University of Chicago, 5640 South Ellis Avenue, Chicago, IL, USA 60637}
\affiliation{Department of Astronomy and Astrophysics, University of Chicago, 5640 South Ellis Avenue, Chicago, IL, USA 60637}
\affiliation{Enrico Fermi Institute, University of Chicago, 5640 South Ellis Avenue, Chicago, IL, USA 60637}
\author{L.~M.~Mocanu}
\affiliation{Kavli Institute for Cosmological Physics, University of Chicago, 5640 South Ellis Avenue, Chicago, IL, USA 60637}
\affiliation{Department of Astronomy and Astrophysics, University of Chicago, 5640 South Ellis Avenue, Chicago, IL, USA 60637}
\author{J.~Montgomery}
\affiliation{Department of Physics, McGill University, 3600 Rue University, Montreal, Quebec H3A 2T8, Canada}
\author{A.~Nadolski}
\affiliation{Astronomy Department, University of Illinois at Urbana-Champaign, 1002 W. Green Street, Urbana, IL 61801, USA}
\affiliation{Department of Physics, University of Illinois Urbana-Champaign, 1110 W. Green Street, Urbana, IL 61801, USA}
\author{T.~Natoli}
\affiliation{Department of Astronomy and Astrophysics, University of Chicago, 5640 South Ellis Avenue, Chicago, IL, USA 60637}
\affiliation{Kavli Institute for Cosmological Physics, University of Chicago, 5640 South Ellis Avenue, Chicago, IL, USA 60637}
\affiliation{Dunlap Institute for Astronomy \& Astrophysics, University of Toronto, 50 St George St, Toronto, ON, M5S 3H4, Canada}
\author{J.~P.~Nibarger}
\affiliation{NIST Quantum Devices Group, 325 Broadway Mailcode 817.03, Boulder, CO, USA 80305}
\author{G.~Noble}
\affiliation{Department of Physics, McGill University, 3600 Rue University, Montreal, Quebec H3A 2T8, Canada}
\author{V.~Novosad}
\affiliation{Materials Sciences Division, Argonne National Laboratory, 9700 S. Cass Avenue, Argonne, IL, USA 60439}
\author{S.~Padin}
\affiliation{Kavli Institute for Cosmological Physics, University of Chicago, 5640 South Ellis Avenue, Chicago, IL, USA 60637}
\affiliation{Department of Astronomy and Astrophysics, University of Chicago, 5640 South Ellis Avenue, Chicago, IL, USA 60637}
\affiliation{California Institute of Technology, MS 249-17, 1216 E. California Blvd., Pasadena, CA, USA 91125}
\author{S.~Patil}
\affiliation{School of Physics, University of Melbourne, Parkville, VIC 3010, Australia}
\author{C.~Pryke}
\affiliation{School of Physics and Astronomy, University of Minnesota, 116 Church Street S.E. Minneapolis, MN, USA 55455}
\author{J.~E.~Ruhl}
\affiliation{Physics Department, Center for Education and Research in Cosmology and Astrophysics, Case Western Reserve University, Cleveland, OH, USA 44106}
\author{B.~R.~Saliwanchik}
\affiliation{Physics Department, Center for Education and Research in Cosmology and Astrophysics, Case Western Reserve University, Cleveland, OH, USA 44106}
\affiliation{Department of Physics, Yale University, P.O. Box 208120, New Haven, CT 06520-8120}
\author{K.~K.~Schaffer}
\affiliation{Kavli Institute for Cosmological Physics, University of Chicago, 5640 South Ellis Avenue, Chicago, IL, USA 60637}
\affiliation{Enrico Fermi Institute, University of Chicago, 5640 South Ellis Avenue, Chicago, IL, USA 60637}
\affiliation{Liberal Arts Department, School of the Art Institute of Chicago, 112 S Michigan Ave, Chicago, IL, USA 60603}
\author{C.~Sievers}
\affiliation{University of Chicago, 5640 South Ellis Avenue, Chicago, IL, USA 60637}
\author{G.~Smecher}
\affiliation{Department of Physics, McGill University, 3600 Rue University, Montreal, Quebec H3A 2T8, Canada}
\affiliation{Three-Speed Logic, Inc., Victoria, B.C., V8S 3Z5, Canada}
\author{A.~A.~Stark}
\affiliation{Harvard-Smithsonian Center for Astrophysics, 60 Garden Street, Cambridge, MA, USA 02138}
\author{C.~Tucker}
\affiliation{Cardiff University, Cardiff CF10 3XQ, United Kingdom}
\author{K.~Vanderlinde}
\affiliation{Dunlap Institute for Astronomy \& Astrophysics, University of Toronto, 50 St George St, Toronto, ON, M5S 3H4, Canada}
\affiliation{Department of Astronomy \& Astrophysics, University of Toronto, 50 St George St, Toronto, ON, M5S 3H4, Canada}
\author{T.~Veach}
\affiliation{Department of Astronomy, University of Maryland College Park, MD, USA 20742}
\author{J.~D.~Vieira}
\affiliation{Astronomy Department, University of Illinois at Urbana-Champaign, 1002 W. Green Street, Urbana, IL 61801, USA}
\affiliation{Department of Physics, University of Illinois Urbana-Champaign, 1110 W. Green Street, Urbana, IL 61801, USA}
\author{G.~Wang}
\affiliation{High Energy Physics Division, Argonne National Laboratory, 9700 S. Cass Avenue, Argonne, IL, USA 60439}
\author{N.~Whitehorn}
\affiliation{Department of Physics and Astronomy, University of California, Los Angeles, CA, USA 90095}
\author{W.~L.~K.~Wu}
\affiliation{Kavli Institute for Cosmological Physics, University of Chicago, 5640 South Ellis Avenue, Chicago, IL, USA 60637}
\author{V.~Yefremenko}
\affiliation{High Energy Physics Division, Argonne National Laboratory, 9700 S. Cass Avenue, Argonne, IL, USA 60439}
\collaboration{SPTpol Collaboration}
\noaffiliation

\begin{abstract}

We report a \bmode{} power spectrum measurement from the cosmic microwave background (CMB) polarization anisotropy observations made using the SPTpol instrument on the South Pole Telescope. 
This work uses 500\,\sqdeg{} of SPTpol data, a five-fold increase over the last SPTpol \bmode{} release.
As a result, the bandpower uncertainties have been reduced by more than a factor of two, and the measurement extends to lower multipoles: $52 < \ell < 2301$.
Data from both 95 and 150\,GHz are used, allowing for three cross-spectra: 95 GHz x 95 GHz, 95 GHz x 150 GHz, and 150 GHz x 150 GHz.
\bmode{} power is detected at very high significance; we find $P(BB < 0) = \positiveBBpte$, corresponding to a $\positiveBBsignif\,\sigma$ detection of power.
With a prior on the galactic dust from Planck, WMAP and BICEP2/Keck observations, the SPTpol \bmode{} data can be used to set an upper limit  on the tensor-to-scalar ratio, $r <  \rlimit$ at 95\% confidence (the expected $1\,\sigma$ constraint on $r$  given the measurement uncertainties is 0.22).
We find the measured \bmode{} power is consistent with the \planck{} best-fit \lcdm{} model predictions.
Scaling the predicted lensing \bmode{} power in this model by a factor \alens{}, the data prefer $\alens =  \abbrzero$.
These data are currently the most precise measurements of \bmode{} power at $\ell > 320$.

\end{abstract}

\keywords{cosmic background radiation -- cosmology: observations}

\maketitle

\section{Introduction}
\label{sec:intro}
\setcounter{footnote}{0}

Measurements of cosmic microwave background (CMB) anisotropy are a cornerstone of modern cosmology.
After recombination at $z \sim 1100$, the overwhelming majority of CMB photons have freely streamed to observers today.
The anisotropy we see primarily arises from fluctuations in the density of the primordial universe during recombination.
Thus, measurements of these photons offer us a snapshot of the universe in its infancy.

The CMB is polarized at approximately the 10\% level, due to Thomson scattering off free electrons illuminated by local radiation quadrupoles.
At $\ell> 10$, polarization is sourced by quadrupole moments that start growing in the primordial plasma and affect the local environment of the photons and electrons as they begin decoupling during recombination.
Being driven by scalar (density) perturbations, the resulting full-sky polarization field has even parity analogous to electric fields, following a gradient-like polarization pattern commonly referred to as ``\emodes".
\emode{} polarization of the CMB has been measured with high precision by e.g., \citet[][hereafter H18]{henning18}, \citet{louis17} and \citet{planck18-5}, adding information to the temperature spectrum \citep{story12,planck18-5} both by approximately doubling the number of modes that can be measured on the sky and extending the measurement to smaller angular scales due to the comparatively lower foreground levels in polarization.

In addition to \emodes,  there are also odd-parity, curl-like polarization pattern components, called ``\bmodes".
An early prediction of inflation was that there would be a stochastic background of gravitational waves on super-horizon scales \citep{guth81}.
Such gravitational waves would imprint a \bmode{} signature on CMB polarization peaking at $\ell<100$.
The search for the inflationary gravitational wave signal in the polarization of the CMB is a matter of intense current interest, as an unambiguous detection would rule out some alternatives to the inflationary paradigm and yield information on what caused inflation by constraining the shape of the inflaton potential. 
 The best current limit on the inflationary gravitational wave, parameterized as the tensor-to-scalar ratio r, is   $r < 0.06$ at 95\% CL \citep{bicep2keck18}, and comes from a combination of data from BICEP2/Keck,  \planck{}, \wmap{},   and other experiments.

Finally, observers today see a distorted version of the primordial map of the CMB radiation at recombination due to the gravitational lensing of CMB photons by large-scale structure.
The small deflections introduced by gravitational lensing do not preserve the even-parity of the initial \emode{} map, and transform a portion of the \emode{} power into so-called ``lensing \bmodes".
The amplitude of the lensing \bmode{} spectrum depends on the integrated gravitational potential, $\phi$, along the line of sight \citep{seljak99}, making it a useful probe of the growth of structure.
In particular, the CMB lensing signal can help constrain the sum of the neutrino masses, as larger rest masses increase the expansion rate and thereby suppress growth \citep{abazajian15b, pan15}.

 The first measurement of lensing \bmodes{} came from cross-correlating the observed \bmodes{} to a template constructed from CMB \emodes{} and a CIB-derived $\phi$ map, as described in \citet{hanson13}.  Since then, direct measurements of the CMB \bmode{} power spectrum have been made by BICEP2/Keck \citep{bicep2keck15,bicep2keck18}, SPTpol \citep{keisler15} (hereafter K15), POLARBEAR \citep{polarbear17, polarbear19} and ACTpol \citep{naess14,louis17}.

The lensing \bmode{} signal, while cosmologically interesting in its own right, is also a contaminating foreground in any search for inflationary gravitational waves.  
Improved measurements of the lensing signal also facilitate ``delensing'' analyses \citep{knox02}, whereby the lensing portion of the \bmode{} signature can be subtracted off, leaving as the residual any potential inflationary gravitational wave \bmode{} signature \citep[e.g.,][]{manzotti17, planck18-8}.

In this work, we present an improved measurement of the \bmode{} power spectrum in the multipole range 52 $\leqslant$ $\ell$ $\leqslant$ 2301 from the SPTpol 500\,deg$^2$ survey.
While the analysis follows the methods in K15 closely, we use five times more sky area in this work (reducing bandpower uncertainties by approximately $\sqrt{5}$), and extend the measurement to lower multipoles in order to constrain the inflationary gravitational wave power as well as measuring lensing \bmodes.

We describe the SPTpol instrument and survey in \S\ref{sec:survey}.
We discuss the reduction of the time-ordered data in \S\ref{sec:datareduc}, the map-making in \S\ref{sec:maps}, and the power spectrum estimator in \S\ref{sec:powspec}.
We test the data for systematic biases in \S\ref{sec:systematics} and then present the resulting bandpowers in \S\ref{sec:bandpowers}.
We discuss the implications for cosmology in \S\ref{sec:interpretation}, and conclude in \S\ref{sec:conclusion}.

\setcounter{footnote}{0}

\section{The SPTpol instrument and survey}
\label{sec:survey}

The South Pole Telescope (SPT) is a 10-meter diameter, off-axis Gregorian telescope located at the Amundsen-Scott South Pole Station \citep{padin08,carlstrom11} that was designed to make high-precision maps of the CMB with arcminute-scale resolution.
The SPTpol instrument replaced the earlier SPT-SZ instrument and was used for observations on the SPT from early 2012 to the end of 2016 (this work uses data through 2015).
SPTpol consists of 1536 polarization-sensitive transition edge sensor (TES) bolometers cooled to 250\,mK; 1176 with bands centered at 150 GHz and 360 at 95 GHz.
Pairs of these bolometers that are fed by a common feedhorn form optical pixels, with each bolometer coupled to orthogonal linear polarizations.
Full information on the detectors can be found in \citet{henning12} and \citet{sayre12}.

This work uses SPTpol observations of a 500\,\sqdeg{} field spanning -50 to -65 degrees in declination and 22h to 2h in right ascension.
The 150\,GHz data was previously used in the \emode{} power spectrum measurement by \citet{henning18}, and we refer the reader to that work for a detailed description of the observing strategy.
Briefly, we observe the field with a series of back-and-forth raster scans at constant declination, following each raster scan by an approximately 9 arcminute declination step until the full declination range has been covered.
Starting declinations are staggered or ``dithered'' between observations to smooth out the coverage pattern of the field.
From April 2013 to May 2014, we observed the field using a ``lead-trail'' scan strategy which split the field in half in right ascension.
From May 2014 onwards, we moved to a full field scan strategy.
With a full field scan, we could increase the scan speed (and shift sky signals of interest to higher frequencies) at the cost of losing the ability to difference the two half-maps to remove any ground contamination.
Note that we find no indication of significant ground contamination and simply combine the lead and trail maps into complete field observations.

\section{Data Reduction}
\label{sec:datareduc}

The data reduction pipeline used for this work is based on the one used by previous SPTpol power spectrum measurements (K15, \citealt[][hereafter C15]{crites15}, H18).
In the following section, we will present a brief overview of the components of the data reduction pipeline, highlighting differences from the procedures used in the works mentioned earlier.

We construct maps using the same procedure as outlined in C15, K15, and H18.
For this work our maps are in the Lambert azimuthal equal-area projection with 1.5 arcminute pixels.
Briefly, the reduced, weighted detector time-ordered-data are bandpass filtered, combined with pointing information, and binned into the appropriate on-sky map pixel.
Binned maps are then cleaned with a second set of cleaning routines before being Fourier transformed and the frequency-domain maps decomposed into the $E$- and \bmode{} basis.
The resulting frequency-domain maps are  binned into pseudo power spectra.

\subsection{Calibration of time-ordered data}
\label{subsec:tod}

The first step is to clean, calibrate and characterize the time-ordered data (TOD). 
The detector TOD have some response to signals seen by near-by detectors due to electrical `cross-talk'. 
Unlike K15 which dealt with cross-talk at the map level, we begin by removing cross-talk between detectors from each detector's time-ordered data. 
This process is described in H18, and is based on building up a decorrelation matrix from individual detector observations of the Galactic \textsc{HII} region RCW38.

Next, we calibrate the individual detector TOD to CMB brightness temperature in a two-step process. 
The first step calibrates single detectors using a combination of the response to an internal chopped calibration source and observations of RCW38. 
This process was first described in \citet{schaffer11} and used in previous SPT results. 
We add a second step because we find that calibrated, pixel-differenced timestreams contain excess residual power at low data frequencies, corresponding to large angular scales on the sky. 
This excess is believed to be due to atmospheric fluctuations. 
To reduce this residual power, we calculate a small correction to each detector's RCW38-derived calibration by finding the scaling factors between detectors in a pixel pair that minimize the low-frequency noise in their differenced timestreams.
For each scan across the field, we calculate the factors $c_{x,y}$  that minimize the pair-differenced noise between 0.1 and 0.3 Hz for every pixel. 
To conserve the mean calibration across the pair, we impose the requirement that $c_x + c_y = 2$ (as $c_{x,y}=1$ represents the initial RCW38 calibration). 
We then take the average $c_{x,y}$ value for each detector across all scans, considering only values where $0.5 < c_{x,y} <1.5$.

It is also crucial to determine the polarization angle of each detector. 
We use the same measured response angles and polarization efficiencies derived for the 100d data in C15 based on measurements of a polarized source 3 km away from the telescope. 
A series of systematic tests, described in both K15 and C15, yield an uncertainty on our per-detector angles of 1\deg, along with a 0.5\deg{} statistical uncertainty from the fits. 
A correlated error in the detector angles will mix power between $E$ and $B$ modes, and is handled by looking at the $EB$ spectrum. 
The mean polarization efficiency is 97\% with a statistical uncertainty of 0.7\%.

Finally, we calculate the weight that the time-ordered data from each detector in an observation should be given when making maps. 
These weights are based on each detector's PSD between 1-3\,Hz.\footnote{The 1-3\,Hz frequency range corresponds roughly to $\ell\in[300,900]$ for a full field observation and $\ell\in[700,2100]$ for a lead-trail observation} 
We difference left-going and right-going scans before calculating this PSD to null any true sky signal, and average the PSD across all pairs of scans in the observation.

\subsection{Time ordered data filtering}
\label{sec:todfiltering}
To reduce the contribution of atmospheric $1/f$ noise to coadded maps, we filter long-wavelength modes, which are expected to be dominated by atmospheric signals in our data, from individual detector time ordered data.
We use as our filtering basis functions Legendre polynomials, up to order 5 for lead-trail observations and order 9 for full-field observations, the same values as in H18.
Each raster scan across the field is filtered over the same range in RA, and the modes removed correspond to spatial modes on the sky at a multipole of less than approximately $\ell$ of 50.

\subsection{Data cuts}

We flag and remove low-quality or pathological data at both the time-ordered data and map levels. 
For instance, these flags remove data from periods when a detector is not properly biased, or when observing conditions drastically reduce sensitivity to sky signals. 
These cuts are summarized below. 

\subsubsection{Time ordered data cuts}
\label{sec:todcuts}
Before binning into maps, we remove data from detectors with corrupted performance as determined by a series of cuts that are very similar to those described in C15 and K15.
There are 96 (249) detectors out of a total of 360 (1176) for the 95 GHz (150 GHz) arrays that fail cuts on each of our 4122 observations.
For the set of remaining ``live'' detectors, we cut those with anomalous performance according to a series of metrics measured from TOD.
The metrics are listed below in the order they are applied to each observation, so a cut detector is counted only by the first test it fails.
The average percentage of detectors at 95 GHz (150 GHz) removed are noted after each cut.

\begin{enumerate}
\item Timestream errors, like a failure to properly bias the detector TES into its transition, readout electronics failure affecting the detector channel, and un-physical calibration of the detector time ordered data into $K_{\rm CMB}$ units; 7.6 (5.8).
\item Anomalously low or un-physically high response to either the chopped internal calibration source or a dip in telescope elevation (which modulates atmospheric loading); 2.1 (0.28)
\item TOD weights, thresholds are empirically set based on distributions of weight values for all observations, which include variability of sky and telescope conditions, to remove un-physical values; 4.3 (3.1).
\item Low-frequency noise, measured between 0.0 and 0.4 Hz in individual detector TOD, calibrated in $K_{\rm CMB}$ units.  Detectors more than 4$\sigma$ away from the mean for all detectors in a given observation are removed; 3.2 (1.6).
\item Broadband noise, measured from the mean power spectral density of all scans for each detector, integrated between 0.4 Hz and 3 Hz, with detectors more than 5$\sigma$ away from the central value being cut; 0.6 (0.8).
\item Full pixel, which removes every bolometer whose pixel partner was cut.  This cut ensures that polarization maps are not corrupted by an uneven sampling of the Q and U maps; 3.6 (1.7).
\end{enumerate}

In addition to the above cuts of a detector's data for a full observation, we flag individual raster scans where a detector experiences a ``glitch'', defined as an anomalous difference ($>10\,\sigma$) between two subsequent data samples.  The distribution of sample-to-sample differences is calculated for the entire focal plane for each scan, and any detector with a difference more than five standard deviations away from the mean has its data for that specific scan cut.

\subsubsection{Cuts on low-frequency map noise}

While we weight the data from individual detectors according to their noise PSDs (see \S\ref{subsec:tod}), this will not necessarily account for observations with unusual levels of correlated noise between detectors, for instance due to atmospheric fluctuations. 
We therefore also implement a cut based on the low-frequency noise in each observation's map. 
We calculate the angular power spectrum of each map's Stokes T, Q, and U components, constructing a metric,  $\Xi_{\alpha}$, defined by:

\bea
\nonumber C^{\alpha,i}_{\ell<300} &=& \sum_{\ell < 300} C^{\alpha,i}_{\ell}\\
  \Xi_{\alpha} &=& \frac{C^{\alpha,i}_{\ell<300}}{\text{median}(C^{\alpha,i}_{\ell<300})}
\eea

where $C^{\alpha,i}_\ell$ is the angular power spectrum of map $i$ and $\alpha$ = (TT, QQ, or UU).  
We cut any map where the low-frequency polarization noise is ten times higher than the median noise, i.e. if either $\Xi_{QQ}$ or $\Xi_{UU}$ is greater than 10. 

After removing maps with anomalously high low-frequency noise, we are left with 3628 good individual half- or full field observations from a total of 4341 performed between March 2013 and November 2015.
Because cuts are applied independently to the lead and trail maps from 2013, sometimes only one of a lead-trail pair passes. 
We combine these orphan half-observations with the nearest-in-time counterpart, and cut the eight half-field maps where a counterpart can not be found.  
We are left with 3620 maps, which, when the lead-trail pairs are combined into observations of the full field, yields a total of 2890 complete observations of the field: 730 in the lead-trail format, 2160 in the full-field format.

\subsubsection{Beams}
\label{sec:beams}
We measure the instrumental angular response function (``beam'') with observations of Venus made in January 2013, that are convolved by an estimate of the effect of pointing uncertainties in the CMB fields. 
A two-dimensional Gaussian is fit to each 1\deg-by-1\deg{} Venus map made with third-order polynomial subtraction, and they are coadded with their best-fit peak pixels aligned.
The resulting two-dimensional map is then convolved with a two-dimensional Gaussian with widths that are determined from fits to a series of bright point sources in the CMB field, measured with nominal pointing information.  
This second step accounts for the ``jitter'' associated with our nominal pointing model.
Thus the convolved Venus map includes the effects of the true instrumental angular response and the variations in pointing over the course of our observations.

The small size of the Venus maps and the use of polynomial filtering of the time-ordered-data mean that the measured beam only has high-fidelity information above $\ell \approx 500$.  
However, as described in H18, we find that the Venus beam profile at large angular scales is in good agreement with an estimate derived using a separate method of cross-spectral analysis between Planck maps and SPTpol maps. 
 As a result, we use the Venus profiles over the full range of multipoles in this analysis.
 We take the variance among our 8 (13) clean Venus maps at 95 GHz (150 GHz) as our beam errors, with the variance among the 8 cross-maps that include our 95 GHz maps as the error on our 95 GHz $\times$ 150 GHz beam.  
 We marginalize over seven beam parameters in the fits, representing the seven largest eigenvectors of the beam covariance matrix. 
 We find our results are robust to doubling the assumed beam uncertainty in Section \ref{sec:results}.

\section{Map Processing}
\label{sec:maps}

We apply further processing at the map level before calculating the power spectrum. 
In particular, we filter out the monopole temperature leakage and signals fixed in RA. 
For computational and coverage reasons, we also bundle together many observations of the field into a bundle map. 
We apodize these maps and mask bright radio sources. 
We then convert the Q/U maps to E/B modes using the $\chi_B$ estimator from \citet{smith07b}. 

\begin{figure*}
\begin{center}
\includegraphics[width=0.70\textwidth, trim=4cm 3cm 4cm 2cm]{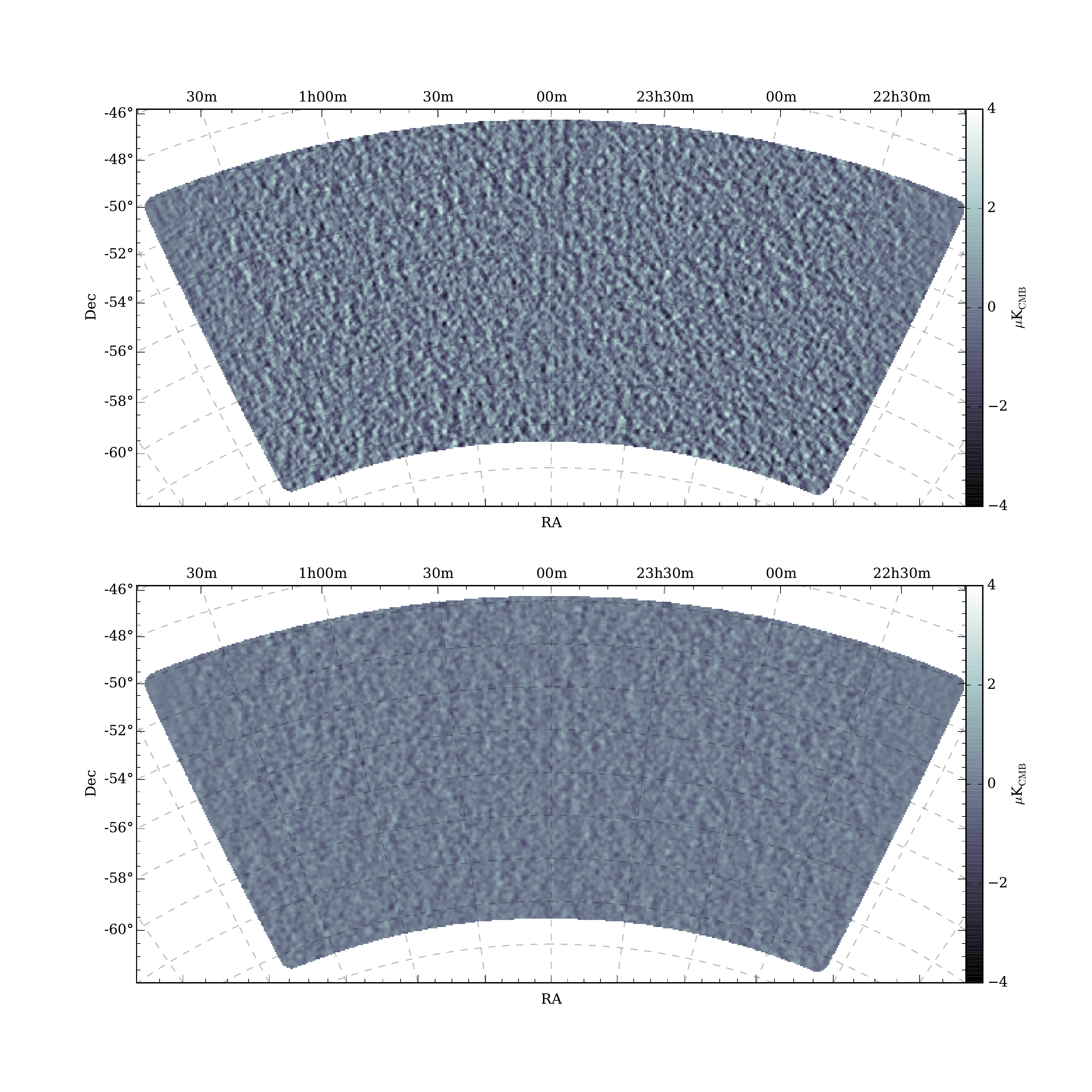}
\end{center}
\caption{The \bmode{} sky maps used in the work, shown as transformed back from frequency-domain maps with all processing steps applied. 
The top panel is 95\,GHz while the bottom panel is 150\,GHz. 
Both maps are noise-dominated on all angular scales. }
\label{fig:b_mode_skies}
\end{figure*}

\subsection{Map bundles}
In order to smooth coverage and reduce the computational demands for later processing steps, we combine the maps into a series of 50 ``bundles''.
For each observation format (lead-trail and full), we combine all constituent maps into a single one and measure its total map weight.  We then divide that total map weight by 50 to get the target per-bundle summed weight.  We then order the maps chronologically (by the start time of the lead observation for mis-matched lead-trail pairs) and combine them sequentially until each bundle is as close as possible to the target per-bundle weight.  The lead-trail-only and full-only bundles are used in the systematics tests described in section \ref{sec:jacks}.  For the final data products, we combine the first lead-trail bundle with the first full bundle and so on until we have 50 bundles, each composed of a lead-trail and full bundle, for both the 95 GHz and 150 GHz data.

\subsection{Apodization and point source masking}
\label{sec:apod_masking}

We apodize the maps before Fourier transforming in order to reduce mode-coupling due to a sharp edge and to downweight the low-weight and high-noise pixels at the edge of the map.  
For simplicity, we use the same apodization mask for all map bundles. 
Thus we begin by finding the intersection across all bundles\footnote{We actually use a larger bundle set, resulting in a smaller intersection region,  to ensure that the mask is also appropriate for all null tests. Namely, we include the individual bundles in the lead-trail vs full and left-going vs right-going splits.}
 of the set of pixels with a weight at least 30\% of the median weight. 
 The combined coverage mask is then reduced by a 4 arcminute border at its edges to reduce edge effects before apodizing the result with a 90 arcminute wide cosine taper.  
The resulting effective sky area after application of the apodization mask is 458.3 square degrees.

While we will marginalize over an unknown Poisson point source power in the parameter fitting, we choose to mask the brightest sources (with intensity fluxes  $> 50$\,mJy at 150\,GHz) to minimize the shot noise. 

\subsection{Map-space processing}

At this point, we have T, Q, and U maps which we want to transform into a \bmode{} map. 
Some systematic sources of apparent \bmodes{} are most readily dealt with in the map domain. 
Thus we project out a temperature map template from each Q/U map and remove a template based on the azimuthal signal before converting from Q/U to E/B maps. 

\subsubsection{Monopole temperature leakage deprojection}
\label{sec:tp_monopole}

Mis-calibrating the gains of two detectors in a pixel causes a scaled copy of the temperature map to leak into the Q and U polarization maps.  
This  ``monopole'' leakage is straightforward to measure and remove in the Q, U maps before they are transformed to $E$ and $B$.
To estimate the leakage, we first construct two half-depth coadds by adding up all even-numbered bundles and all odd-numbered bundles.
The resulting maps are crossed to produce TQ and TU pseudo-cross-spectra, which are each normalized by the TT pseudo-cross-spectrum.
Normalized cross-spectral ratios are then averaged over a chosen ell range, $\ell$ = 100-3000 in this work, yielding coefficients, $\hat{Q}$ and $\hat{U}$, of the T-to-P monopole leakage.
The coefficients, $\hat{Q}$ = 0.0263 (0.0162), $\hat{U}$ = -0.0215 (0.0095) for 95 (150) GHz maps, are insensitive to the exact choice of ell range, but we choose the range where our expected cosmological signal is maximal.
Each bundle Q and U map then has the appropriately scaled version of its own T map subtracted from it.

\subsubsection{RA template removal}
We find evidence for scan-synchronous signals in our bundle maps, with an rms of approximately 4 (1.5) $\mu K$ in 95 (150) GHz Q and U maps, similar in scale to the signals in the 100d field, described in K15.
To remove it, we measure a one-dimensional profile by binning bundle map pixels by their RA location and smoothing the result by a 1-degree wide Hann window.
The resulting profile is then re-projected into two dimensions along the elevation direction and subtracted from the bundle Q and U maps.

\subsection{$E$ and $B$ mode maps}
After applying the real-space processing steps described above to our T, Q, and U maps, we decompose them into harmonic space $T$, $E$, and $B$ maps for further processing and power spectrum estimation.
We construct the \emode{} maps with the standard transformation \citep{zaldarriaga01},
\be
  E_\ell=Q_\ell \cos (2\phi_\ell) + U_\ell \sin (2\phi_\ell)
\ee
where $Q_\ell$ and $U_\ell$ are Fourier transforms of the processed and apodized real-space Q and U maps, and $\phi_\ell=\arctan(\ell_x/\ell_y)$.
A generic effect of the $E$-$B$ decomposition with partial sky coverage is the presence of ambiguous modes, which mix \emode{} sky signal into the constructed \bmode{} map.
To minimize this effect, we use the $\chi_B$ estimator from \citet{smith07b}.
Our final Fourier-space \bmode{} maps are thus constructed according to:
\be
  B_\ell = \frac{\mathcal{F}(W((\partial_X^2 -\partial_Y^2)Q + 2 \partial_{XY}^2 U))}{\sqrt{\alpha_\ell}}
\ee
where $\mathcal{F}$ represents the Fourier transform, $\alpha_\ell = l(l-1)(l+1)(l+2)$,  W is the apodization mask and the derivatives of Q and U  are the intermediate $\chi_B$ maps.
The derivatives are calculated using finite differences with a 5 x 5 pixel kernel centered on each map pixel.
The resulting $B$-mode maps are shown in Fig.~\ref{fig:b_mode_skies}.

\section{Power spectrum estimation}
\label{sec:powspec}

We estimate the \bmode{} bandpowers using a pseudo-C$_\ell$ cross-spectrum method \citep[see K15,][]{hivon02,tristram05}. 
Starting from the cleaned Fourier-space \bmode{} maps of the last section, we average across the set of all cross-spectra to measure the pseudo-spectrum. 
We then correct this  pseudo-spectrum for effects such as the finite sky coverage to create an unbiased estimate of the true \bmode{} power on the sky.

The binned pseudo-C$_\ell$ spectrum is calculated from the mean of the cross-spectra between all bundle map pairings:
\be
   \hat{D}_b^{x\times y} = \frac{1}{N_{x\times y}}\sum_{i\ne j} \sum_{\ell \in b}\left( \frac{\ell(\ell+1)}{2 \pi} Re\left(W_\ell(B_\ell^{x,i} B_\ell^{y,j*})\right) \right).
\ee
Here x and y denote 95 or 150\,GHz, and i or j denote the bundle number. 
The Fourier-space \bmode{} map for frequency x and bundle i is $B_\ell^{x,i}$, while $W_\ell$ is a Wiener-filter-derived mode weighting.  
$N_{x\times y}$ is the number of cross-spectra: there are a total of 1225 cross spectra for the 90 and 150\,GHz auto-spectra and 2450 for the 95 GHz x 150 GHz spectrum.

This binned pseudo-spectrum $\hat{D_b}$ is related to the true binned spectrum $D_{b}$ by:
\be
\hat{D_b}=K_{bb'}D_{b'} + A_b +A_{TB}+A_{EB}.
\ee
We refer to these binned spectra as bandpowers. 
Here $A_b$ captures additive biases to the \bmode{} power created for instance by the map filtering, while $A_{TB}$ is to allow for effects  such as very low amplitude polarised beam sidelobes that are not in the simulations. 
In principle, $A_b$ should be written as a function of \{$C_\ell^{TT}, C_\ell^{EE}, C_\ell^{TE}$\} but the temperature and \emode{} power spectra have already been measured to high precision so we fix $A_b$ to the expectation for the fiducial cosmology. 
We remove $A_{TB}$ and $A_{EB}$ by subtracting 
\be \label{eqn:tb}
D^{BB}_{b'} =D^{\prime BB}_{b'} - \frac{(D^{TB}_{b'})^2}{D^{TT}_{b'}}- \frac{(D^{EB}_{b'})^2}{D^{EE}_{b'}}.
\ee
In principle, the $A_{EB}$ term could be introduced by an mis-calibration of the polarization angles, as discussed in \S\ref{subsec:tod}. 
In practice, it is very close to zero suggesting that the fiducial polarization angles are accurate. 
The maximum value of this term for a 150\,GHz bandpower is 0.001\,\muksq{}. 
Note that we handle each spectrum (e.g., TT, BB) independently for both simulations and real data, and do not include off-diagonal blocks such as (TT,BB) in the mode-coupling matrices. 
The kernel matrix $K_{bb'}$ encapsulates the effects of mode-mixing due to partial sky coverage, and the suppression of power by the instrumental beam and map filtering.

\subsection{Estimating the additive biases}
\label{subsec:additive}

We measure the induced additive bias in the \bmode{} power spectrum by measuring the observed \bmode{} power in a suite of 100 $TE$-only simulations (see \S\ref{sec:sims}). 
The additive bias can be understood by considering the ambiguous modes that are created by the interaction of the partial sky coverage and edge apodization with the polynomial filtering applied to the TOD. 
These ambiguous modes mix the \emode{} power into \bmodes, particularly at low angular multipoles. 
Because the \emode{} power spectrum is tightly constrained and we can accurately simulate the TOD processing, we can determine the expectation value for the additive bias, $A_b$,  using the 200 $TE$ simulations. 
As with the real data, we first subtract the $T$-$B$ leakage estimate (Eqn.~\ref{eqn:tb}) from each individual $TE$ simulation. 
In the lowest multipole bin, the additive bias is larger than the expected \bmode{} power, quickly falling and becoming negligible at higher angular multipoles. 
Specifically, the additive bias is $\sim$35\% of the expected power in second bin ($\ell\in[152,301]$), $\sim$10\% in the third bin, and about 3\% at higher multipoles.  
We also subtract the estimated $A_b$ from the $TEB$ simulations used for estimating the bandpower covariance, with the variations about the mean additive bias adding to the sample variance estimate.

\subsection{Estimating the kernel matrix}
\label{sec:powspec_debias}

We also need to calculate the kernel matrix $K_{bb'}$ in order to apply its inverse and recover an unbiased estimate of the true sky power spectrum. 
The kernel matrix, which includes the effects of binning, TOD filtering and map-making, instrumental beams, and mode mixing due to edge apodization and finite sky coverage, can be written as:
\be
K_{bb'}=P_{b\ell'}\left(M_{\ell \ell'}T_{\ell'}\mathcal{B}_{\ell'}^2\right)Q_{\ell'b'}.
\ee
Here $P_{b\ell}$ and $Q_{\ell b}$ are the binning and un-binning operators \citep{hivon02} that translate between bandpower-space and native $\ell$ space, and $M_{\ell\ell}$ is the mode-mixing matrix that describes the $\ell$-space mixing induced by finite sky coverage and edge apodization of our field. 
In this work, given the relative lack of features in the \bmode{} power spectrum at the current signal-to-noise, we make the simplifying approximation that the mode-mixing matrix is diagonal. 
The measurement of the azimuthally-averaged beam $\mathcal{B}_\ell$ was described in \S\ref{sec:beams}, and $T_{\ell}$ is the filter transfer function described next.

\subsubsection{Filter transfer function}
\label{sec:tranfunc}

We estimate the effect of the TOD filtering and the map-making process using a set of noiseless simulated CMB skies. 
Each sky realization is passed through the full TOD processing, map-making and conversion from Q/U to Wiener-filtered $E$$B$ maps. 
The $BB$ power spectrum is then calculated for each of the 100 $TE$ and 100 $TEB$ skies (see \S\ref{sec:sims}). 
The $TE$ skies are required to estimate any additive bias to the \bmode{} spectrum. 
As with the real data, we subtract the $T$-$B$ leakage estimate (Eqn.~\ref{eqn:tb}) from each individual $TE$ and $TEB$ simulation. 
We subtract the mean $TE$ \bmode{} spectrum from the mean $TEB$ \bmode{} spectrum. 
As we have assumed a diagonal mode-mixing matrix, we simply take the ratio of this cleaned spectrum to the product of the beam function $\mathcal{B}_\ell^2$ and combined CMB and foreground spectrum  $C_\ell$ as the one-dimensional transfer function $T_\ell$.

\subsection{Bandpower covariance}
\label{sec:bandpower-cov}

The bandpower covariance includes contributions from sample and noise variance. 
The noise variance is estimated through the covariance between individual cross-spectrum realizations from our ensemble of noise-only bundles, while the sample variance is calculated from the scatter in the set of autospectra of the simulated signal-only $TEB$ map realizations described in \S\ref{sec:sims}. 
Note that these simulations are for $r=0$.  
We combine these two estimates together according to:
\bea
\nonumber \Xi_{bb}^{XX} &=& 2 \gamma_b  ( S_b^{XX} +N_b^{XX})^2 \\
\Xi_{bb}^{XY} &=& \gamma_b  [ (S_b^{XY})^2 + (S_b^{XX}+N_b^{XX})(S_b^{YY}+N_b^{YY}) ], 
\eea
where $b$ denotes the multipole bin, $X$/$Y$ represent either 95 or 150\,GHz, and $S$ and $N$ are the signal and noise power respectively. 
The prefactor, $\gamma_b$, accounts for the number of modes in each multipole bin. 
For the noise terms (both $N^2$ and $SN$), we make the simplifying assumption that the bandpower covariance matrix between two different multipole bins $b \ne b^\prime$ is zero, i.e.~the matrix is block diagonal. 
While we expect a small degree of correlation between neighboring bins due to the finite sky coverage, this correlation is minimized because the chosen bin sizes are significantly wider than the expected angular multipole resolution for the field size. 
Thus the noise variance should be approximately diagonal. 
As a check on this assumption, we compute the $\chi^2$ statistic of our bandpowers relative to a fiducial cosmology spectrum with both the block-diagonal covariance and the version preserving the noisy estimates of the off-diagonal structure. 
 We find a $\Delta \chi^2$ of approximately 1 for the 21 bandpowers. 
 We allow the sample variance terms ($S^2$) to have an arbitrary shape since the lensing-induced \bmodes{} should have correlations between multipoles. 
We also stress that  bin-bin correlations are included for the beam and calibration uncertainties, which are dealt with separately.

We add to the covariance an estimate of the 1\,$\sigma$ variations in the T-to-P monopole leakage terms in \S\ref{sec:tp_monopole}.

\subsection{Power spectrum calibration}
\label{sec:powspec_cal}
As described in section \ref{subsec:tod}, we initially calibrate our detector data in units of CMB brightness temperature by fitting to a known-brightness source.
We further refine our calibration by cross-correlating SPTpol $T$- and \emode{} maps to the published Planck maps over our nominal observation region, as described in H18.
Because we use nearly identical data sets and processing options as in H18, we take the median  of that work's calibration posterior  as our 150 GHz polarization calibration factor, $P_{\text{cal}}^{150}$. 
The uncertainty on $P_{\text{cal}}^{150}$ is 0.5\% based on the H18 posterior.

As H18 only used 150\,GHz data, we must also extend the known 150\,GHz polarization calibration to 95\,GHz. 
We do this  by constructing an ensemble of ratio spectra between the 95x150 GHz and 150x150 GHz pseudo cross spectra:

\be
\epsilon_{\ell,i} =\frac{C_{\ell~EE,i}^{~95\times150}}{C_{\ell~EE,i}^{~150\times150}} \frac{\mathcal{B}_\ell^{150}}{\mathcal{B}_\ell^{95}}.
\ee
Here $i$ denotes which cross-spectrum. 
We average each ratio spectrum over the $\ell$-bins  with inverse variance weighting to yield an ensemble of ratio factors, $\epsilon_i$, and take as our 95 GHz polarization calibration scaling the value $P_{\text{cal}}^{95}=<P_{\text{cal}}^{150}><\epsilon_i>$.
We estimate the uncertainty in $<\epsilon_i>$ from the spread in the $\epsilon_i$ ensemble, and estimate a combined uncertainty of 5.2\% in the 95\,GHz calibration by adding the two uncertainty terms in quadrature. 
We have also confirmed that the measured E-mode power spectra from these maps at 95 and 150\,GHz are consistent with the reported bandpowers in H18. 
We marginalize over two calibration parameters in all fits, representing the 95 and 150\,GHz calibration factors, with priors set from the above calculation.


\begin{figure*}

\begin{center}
\includegraphics[width=0.90\textwidth, trim =  2.2cm  13cm  2cm 3cm]{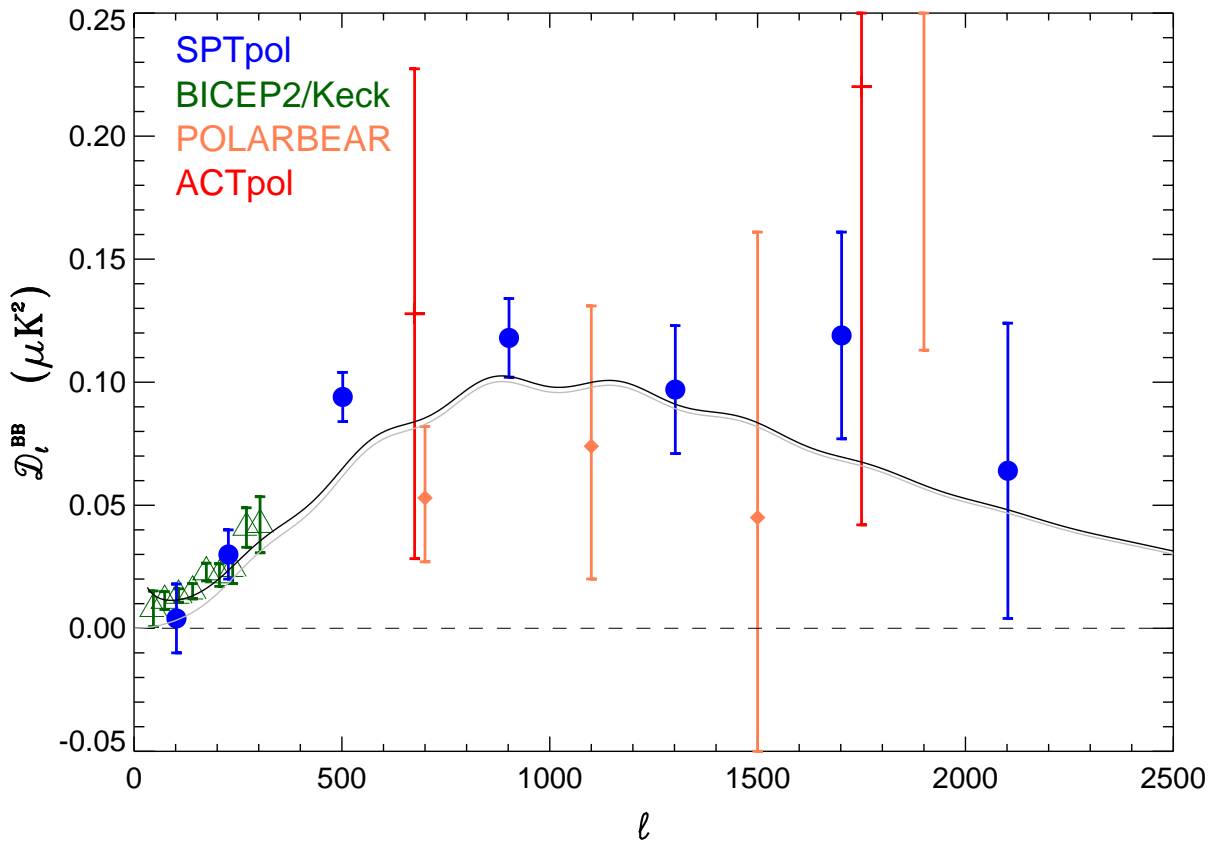}
\end{center}
\caption{
\label{fig:BB_spectra}
The minimum-variance $B$-mode bandpowers from this work (blue circles) along with the $B$-mode measurements from other experiments. 
The 150\,GHz results for BICEP2/Keck \citep{bicep2keck18} are shown by the green triangles. 
ACTpol results at 150\,GHz \citep{louis17} are marked by the red crosses, and POLARBEAR measurements at 150\,GHz \citep{polarbear17} by the orange diamonds. 
The grey line shows the prediction for lensed $B$-mode power for the \textsc{Planck} best-fit model, while the black line adds on the best-fit Galactic dust power from \citet{bicep2keck18}. 
The $B$-mode bandpowers from this work  are the most precise measurements at $\ell > 320$.
}
\end{figure*}

\subsection{Simulations}
\label{sec:sims}

A crucial element of our power spectrum analysis is the use of simulated skies.
We start with 100 Lenspix realizations of lensed $T$, $E$, $B$ skies, generated from the \planck{} + WP + high L cosmology in \citet{planck15-13}, and add in Gaussian foregrounds realizations.
The foreground terms include polarized Galactic dust with a polarized power of 0.0236\,\muksq{} at $\ell=80$ and 150\,GHz; unpolarised thermal Sunyaev-Zel'dovich effect power with an amplitude of 5\,\muksq{} at $\ell=3000$ and 150\,GHz; Poisson dusty star-forming galaxies at a power level of  9\,\muksq{} at $\ell=3000$ and 150\,GHz and polarization fraction of 0.025; Poisson radio galaxies at a power level of  10\,\muksq{} at $\ell=3000$ and 150\,GHz and polarization fraction of 0.025; and lastly clustered dusty star-forming galaxies with a power level of 5\,\muksq{} at $\ell=3000$ and 150\,GHz. 
The CMB and foreground $a_{\ell m}$'s are combined and convolved with a temporary detector beam profile.
The beam used in simulations is an approximation of the production beam described in section \ref{sec:beams}, as the beam analysis was not complete when simulations were generated.\footnote{The differences between the final and simulation beams are small, typically at the subpercent level with a maximum fractional difference of a few percent.}
At this stage, we make a copy of each set of beam-convolved $a_{\ell m}$'s and zero its \bmodes, allowing us to track the leakage of power from (T, $E$) into $B$ due to our map processing steps.
The $T$, $E$, and $B$ $a_{\ell m}$'s are then converted back into T, Q, and U according to our field definition and projected onto a grid of cylindrical coordinates at twice the resolution of our final maps.
We then mock-observe the T, Q, and U skies to produce pairs of noiseless $TEB$ sky and associated $TE$-only sky maps.
The T maps are identical between the two sets, while the Q and U maps differ only by the lack of source $B$ modes in the latter set.

\Comment{
\begin{table*}
\begin{center}
\caption{Simulation foreground parameters}
\begin{tabular}{c | c | c | c | c | c | c | c | c | c }
  \hline
  \hline
  Source    & $A_S^{\text{~DUST}}$ &  $A_S^{\text{~RADIO}}$ & $A_{~\text{CIB}}$ & $A_{\text{~TSZ}}$ & $P_{\text{~CIB}}$ & $A_{\text{~DUST}}^{EE,80}$ & $\rho_{\text{~POISSON}}$ & $\rho_{\text{~DUSTY}}$ & $\rho_{\text{~RADIO}}$  \\
  95 GHz   & 1.5                     &       50.                 &    0.56        &     12.0        &     0.8         &        0.00338               &       0.025                 &    0.0004             &       0.0014            \\
  150 GHz &      9.                  &             10.           &      5.0        &      5.0         &      0.8        &     0.0236                    &      0.025                  &       0.0004          &       0.0014            \\
   \hline
  \end{tabular}
  \label{tab:sim_foregrounds}
\end{center}
\end{table*}}

\section{Systematic tests}
\label{sec:systematics}

We now turn our attention to potential sources of systematic error in the reported bandpowers. 
First, we look at a suite of null tests to validate the bandpowers against unexpected systematics. 
Then, we examine the sensitivity of the power spectrum to possible systematics which were not tested by the null test suite. 
We find the BB bandpowers are not significantly impacted by systematic biases.

\subsection{Null tests}
\label{sec:jacks}
To check for systematic contamination in our data, we create difference or null maps that will null the true sky signal while maximizing the potential systematic signal for various potential systematics. 
For each potential systematic, we start from a set of 100 maps, order the maps according to the relevant statistic, and then difference the first 50 from the second 50 maps to create 50 null maps. 
We calculate the bandpowers for this 50 null maps, which should be consistent with zero in the absence of systematics. 
\begin{enumerate}
  \item Azimuth: The CMB field rotates relative to the ground throughout the course of observations, so by bundling maps according to the azimuth angle of the ground under the field during observations, we can isolate contamination from sources at the South Pole station.
  \item Lead-Trail/Full: In addition to changing the raster pattern, which in turn affects the weights of full-field maps, the switch from lead-trail to full observations included increasing the scan speed.  Each jackknife bundle consists of a lead-trail bundle differenced with a combined bundle of full observations taken at a similar time in subsequent years.
  \item Moon: Jackknife maps are constructed by combining bundles, without respect to lead-trail or full observation strategy, according to the nearness of the moon to the observation field.
  \item Sun: Jackknife maps are constructed similarly to the moon jackknife maps, except using the presence of the sun in the sky, which occurs at both the very beginning and end of each observing season.
  \item Left -- Right: Each of the Lead-Trail/Full bundles is constructed from maps consisting of rightgoing-only and leftgoing-only scans, with the two sets of scans usually combined to get a full coverage observations.  For this jackknife test, the rightgoing-only and leftgoing-only bundles are differenced, to test for scan-synchronous contamination that depends on the direction of telescope motion.  In particular, rightgoing scans always follow an elevation step, so any ``wobble'' in the telescope due to the elevation motion would stand out in a right - left difference.
\end{enumerate}

We calculate the probability-to-exceed (PTE) of the $\chi^2$ values of each set of jackknife bandpowers for each frequency combination (95x95, 150x150, and 95x150) and each sky combination ($B$x$B$, $E$x$B$, and $T$x$B$) relative to a null spectrum.  The individual jackknife test PTEs support the case of no contamination, with only two of the 45 PTEs outside the interval of (0.05, 0.95).  As a further distillation of the $\chi^2$ information, the PTEs relative to null of the combined bandpowers for all tests and frequency combinations are 0.67 (BxB), 0.20 (ExB), and 0.19 ($T$x$B$), and 0.38 for all bandpowers across all spectra and frequency combinations.

In addition to the basic $\chi^2$ PTE tests, we repeat the process described in K15, whereby individual ``$\chi$ bandpowers'', defined as
 \be
  \chi_b^{fsj}\equiv \frac{C_b^{fsj}}{\sigma \left ( C_b^{fsj} \right)}
 \ee
 are compared to 100000 simulated ensembles generated from unit-width, zero-mean Gaussian distributions. 
The superscripts represent the spectrum  ($f \in \{BB, EB, TB\}$), frequency combination ($s \in \{$95$\times$95$, $95$\times$150$, $150$\times$150$\}$), and null test ($j$), while the subscript $b$ represents a specific $\ell$ bin. 
  We construct a series of test statistics that probe various potential signatures of systematic contamination, summarized in Table \ref{tab:chitests}, and measure how often the statistic as calculated from the simulated $\chi$ bandpowers exceeds the value from a particular jackknife.

\begin{table}[h]
\begin{center}
\caption{Null test PTEs}
\begin{tabular}{l | c }
   Test Statistic & PTE \\
   ${\rm max}_{fsj}\left(|\Sigma_b\chi_b^{fsj}|\right)$ & 0.24\\
   ${\rm max}_{fsj}\left((\chi_b^{fsj})^2\right)$ & 0.60\\
   ${\rm max}_{fsj}\left(\Sigma_b (\chi_b^{fsj})^2\right)$ & 0.92 \\
   $\Sigma_{bfsj}(\chi_b^{fsj})^2$ & 0.13\\
   Global & 0.38
 \end{tabular}
 \label{tab:chitests}
\begin{tablenotes}
The PTEs listed indicate how often statistic in question is higher in the simulated $\chi$ bandpowers than in our ensemble of jackknife $\chi$ bandpowers.  The global PTE indicates how often all four statistics in from a realized ensemble of $\chi$ bandpowers exceed the values from our jackknife bandpowers. 
The top row ($max_{fsj}\left(|\Sigma_b\chi_b^{fsj}|\right)$) tests for spectra that are preferentially positive or negative. 
The second row ($max_{fsj}\left((\chi_b^{fsj})^2\right)$) tests for individual outliers. 
The third row ($max_{fsj}\left(\Sigma_b (\chi_b^{fsj})^2\right)$) is sensitive to null spectra with a larger than expected number of outliers. 
The fourth row is the total $\chi^2$ for all spectra, frequency combinations and null tests. 
The last row, ``Global", is the fraction of simulations that have larger PTEs for all four tests simultaneously. 
\end{tablenotes}
\end{center}
\end{table}

We take the values in Table \ref{tab:chitests}, summarized by the global $\chi^2$ PTE as strong evidence that our data is not contaminated by any of the potentials sources of systematic contamination we investigated.

\begin{figure*}[htb]
\begin{center}
\includegraphics[width=0.9\textwidth, trim =  1.85cm  12.06cm  9.20cm 10.21cm]{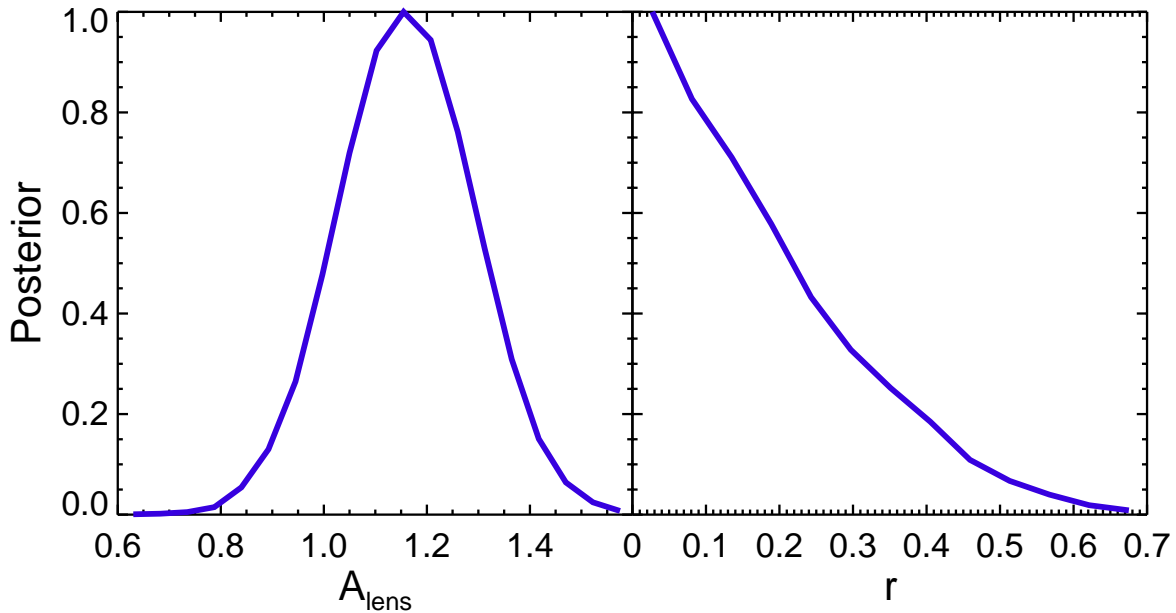}
\end{center}
\caption{The measured \bmode{} power spectrum is consistent with the \planck{} best-fit \lcdm{} model.
On the left, we show the posterior probability for \alens{} (\alens{} rescales the predicted \bmodes{} due to lensing), finding it consistent with unity.
In the right panel, we show that the posterior probability for the tensor-to-scalar ratio r peaks at zero.
}
\label{fig:param}
\end{figure*}

\subsection{Other possible systematics}

Now we turn to two systematics that are not tested by the jackknives.
In contrast to K15, we remove crosstalk directly from time ordered data before binning into maps.
We also explicitely remove monopole $T \rightarrow Q\/U$ leakage from maps before transforming into harmonic space.
Thus, the dominant sources of leakage are expected to be \EtoB{} leakage from filtering. 
These leakage terms are accounted for in \S\ref{subsec:additive} using the observed $B$-modes after filtering the $TE$-only sims, with variance in the leaked power showing up as additional sample variance.

Variations in detector responsivity as a function of the observing elevation would not be detected by jackknife tests.
To probe this we generate half-map masks, starting from the non-apodized mask described in \S\ref{sec:apod_masking} and zeroing either the portion greater or less than declination of -57.5$\deg$.
The resulting masks are then apodized with the same parameters as the real data mask and used to estimate two sets of power spectra. 
We find no evidence of inconsistency between either set of half-map spectra and the full-map spectra,  with a $\chi^2$ PTE of 0.21 when comparing the sub-field bandpowers and their diagonal covariances to the full-field bandpowers.

\section{Bandpowers}
\label{sec:bandpowers}

\begin{figure}
\begin{center}
\includegraphics[width=0.5\textwidth]{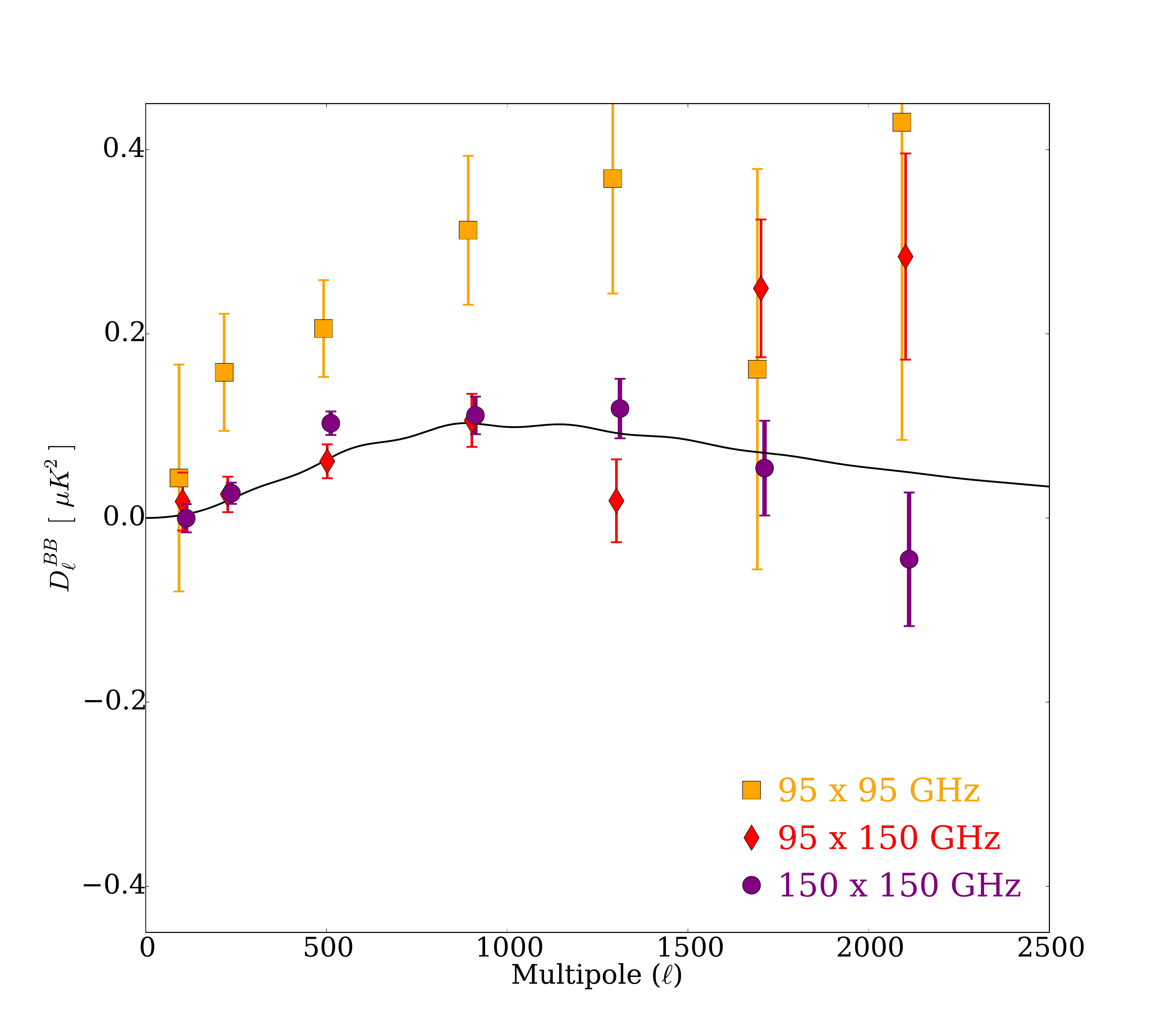}
\end{center}
\caption{\label{fig:SPTpol_bandpowers}
The $BB$ power spectrum bandpowers from the individual
95\,$\times$95\,GHz (orange squares),
95\,$\times$150\,GHz (red diamonds), and
150\,$\times$150\,GHz (purple circles) spectra. 
For reference, the expected lensed $BB$ spectrum from the \textsc{Planck+lensing+WP+highL} best-fit model from \cite{planck13-16}, also used as the source spectrum for simulated skies, is shown as the black solid line.
}
\end{figure}

The final minimum-variance combination of the debiased bandpowers is compared to the results from other experiments in Figure \ref{fig:BB_spectra}. 
The bandpowers for each frequency combination are provided in Table \ref{tab:SPTpol_bandpowers} and plotted in Figure \ref{fig:SPTpol_bandpowers}.  Above $\ell=300$, the bandpower definitions are identical to K15, and the two sets of spectra are consistent, with the uncertainties in this work reflecting the expected $\sqrt{5}$ reduction in total variance from greater sky coverage.  Along with the bandpowers, both figures show the same \planck{} best-fit cosmology described in \ref{sec:results}.


\begin{table*}[ht!]
\caption{BB bandpowers, $D_\ell$ [$ \mu$K$^2$]}
\label{tab:SPTpol_bandpowers}
\vspace{-0.2in}
\begin{center}
\begin{tabular}{ c c  | r  r |  r  r | r  r | r  r}
\hline
\hline
&&\multicolumn{2}{c}{$95 \times 95$}
&\multicolumn{2}{c}{$95 \times 150$}
& \multicolumn{2}{c}{$150 \times 150$}
& \multicolumn{2}{c}{Combined} \\

$\ell_{\rm{center}}$ & $\ell$ range
& $D_\ell$ & $\sigma (D_\ell)$
& $D_\ell$ & $\sigma (D_\ell)$
& $D_\ell$ & $\sigma (D_\ell)$
& $D_\ell$ & $\sigma (D_\ell)$  \\
\hline
102 & 52-151 & 0.043 & 0.123 & 0.018 & 0.032 & -0.000 & 0.015 & 0.004 & 0.014 \\
227 & 152-301 & 0.158 & 0.064 & 0.026 & 0.019 & 0.027 & 0.011 & 0.030 & 0.010 \\
502 & 302-701 & 0.206 & 0.053 & 0.062 & 0.018 & 0.103 & 0.013 & 0.094 & 0.010 \\
902 & 702-1101 & 0.313 & 0.081 & 0.106 & 0.029 & 0.111 & 0.020 & 0.118 & 0.016 \\
1302 & 1102-1501 & 0.369 & 0.125 & 0.019 & 0.045 & 0.119 & 0.032 & 0.097 & 0.026 \\
1702 & 1502-1901 & 0.162 & 0.217 & 0.249 & 0.075 & 0.054 & 0.051 & 0.119 & 0.042 \\
2102 & 1902-2301 & 0.430 & 0.345 & 0.284 & 0.112 & -0.045 & 0.073 & 0.064 & 0.060 \\
\hline
\end{tabular}
\begin{tablenotes}
The BB bandpowers.
\end{tablenotes}
\end{center}
\end{table*}

\section{Interpretation}
\label{sec:interpretation}

We now look at the consistency of the bandpowers with the \lcdm{} model.
While the bandpowers in this work are the best measurements of the \bmode{} power spectrum above $\ell > 320$, we do not expect them to substantially restrict the allowed parameter space within the \lcdm{} framework.
However, these data are interesting as an independent consistency test of the \lcdm{} framework and in the implications for inflationary gravitational waves.

\subsection{Parameter fitting}
\label{sec:param_fitting}
We use the Markov Chain Monte Carlo (MCMC) package \textsc{CosmoMC} \citep{lewis02b} to fit the bandpowers to a simple model of the form:
\be
D_\ell^{\nu_1\times\nu_2} = r D_\ell^{\rm tens}(r=1) + A_{\rm lens} D_\ell^{\rm lens} + D_\ell^{{\rm fg};\, \nu_1\times \nu_2}
\label{eq_bandpower_model}
\ee
We calculate the two templates $D_\ell^{\rm tens}$ and $D_\ell^{\rm lens}$ using CAMB \citep{lewis99} at a \planck{} best-fit cosmology: \{$\Omega_bh^2 = 0.022294, \Omega_ch^2 = 0.11837, \theta = 1.041042, \tau = 0.0677, \text{log}A = 3.0659, n_s = 0.969$\}, with $\sum m_\nu = 60$\,meV. 
Details on how to install and use the SPTpol likelihood and dataset are available on the SPT website.\footnote{http://pole.uchicago.edu/public/data/sayre19/}

The foreground terms included are Galactic dust emission at large angular scales and Poisson power due to polarized emission from extragalactic galaxies at small scales.
We have clear predictions for the Galactic dust emission from \planck{} and BICEP2/Keck; there are only upper limits on the Poisson power as of yet.
The functional form of these foreground terms $D_\ell^{fg;\, \nu_1\times \nu_2}$ for the $\nu_1 \times \nu_2$ cross-spectra is:
\bea
D_\ell^{fg; \,\nu_1\times \nu_2} &=& A_{\ell=80; 150\,GHz}^{\rm dust} f_{\nu_1, \nu_2} \left(\frac{\ell}{80}\right)^{-0.58}  \nonumber\\
&&+ A_{\ell=3000;\,\nu_1\times \nu_2}^{\rm Pois} \left(\frac{\ell}{3000}\right)^2
\eea
Here the top line has the expression for Galactic dust and the bottom line the expression for the extragalactic power.
The frequency dependence of the Galactic dust is encoded in $f_{\nu_1, \nu_2}$ which we assume to be a grey-body spectrum with temperature $T=19.6\,K$ and $\beta=1.59$ \citep{planck13-22}.
Motivated by the recent measurements by BICEP2/Keck on the same region of sky \citep{bicep2keck18}, we place a Gaussian prior on $A_{\ell=80; {\rm 150\,GHz}}^{\rm dust} = 0.0094 \pm 0.0021\, \muksq$. 
We also take the angular shape (i.e.~$D_\ell \propto \ell^{-0.58}$) from the best-fit in that work.  
With only one bin across the relevant angular scales, we do not independently constrain the angular shape of the Milky Way's emission. 
\citet{bicep2keck18} also show that galactic synchrotron is negligible on this field at 95 or 150\,GHz for the current uncertainties.
We make no assumptions about the spectral dependence of the extragalactic Poisson power and thus have three parameters $A_{\ell=3000; \nu_1, \nu_2}^{\rm Pois}$ describing the Poisson power at $\ell=3000$ in the $\nu_1 \times \nu_2$ bandpowers.
We note that of these foreground terms, the data only shows a significant preference for the 95\,GHz Poisson power ($\Delta\chi^2 = 11.4$ for 1 d.o.f.); the others are included to estimate realistic uncertainties.

\subsection{Results}
\label{sec:results}

The \sptpol{} bandpowers are visually consistent with the \planck{} best-fit \lcdm{} cosmology with $r=0$.
Assuming $r=0$, we find $\alens{} = \abbrzero$ which is consistent with the expected value of unity.
Allowing $r$ to vary as well does not significantly change the \alens{} constraints.
The 95\% CL upper limit on $r$ from the SPTpol BB bandpowers only is $r < \rlimit$.
The likelihood curves for \alens{} and $r$ are shown in the left and right panels of Fig.~\ref{fig:param} respectively. 
We also calculate the goodness of fit of the model with $r=0$\footnote{Allowing $r$ to vary does not improve the fit quality significantly.} to the data; the PTE is low at 2\%. 
This is driven by the three $95 \times 150$\,GHz bandpowers above $\ell=1102$.

Both of these results are robust against the assumed Galactic dust prior.
Increasing (or decreasing) the central value of the prior on the Galactic dust power by 50\% only slightly decreases (increases) the central value to $\alens{} = \abbhifg{}$ (\abblofg{}).
The resulting $r$ limits go to $r<\rlimithifg$ and $r<\rlimitlofg$ respectively. 
Removing the external dust prior altogether minimally changes the result to $r<\rlimitnogalprior$.
Thus the results do not depend closely on the details of the dust prior.

The results are driven primarily by the 150x150\,GHz bandpowers, and the 95x95\,GHz bandpowers have little weight in the parameter fits.
We have confirmed this by removing each frequency combination, 95x95\,GHz, 95x150\,GHz, or 150x150\,GHz, and re-running the MCMCs.
Without the 95x95\,GHz bandpowers (which appear high), the recovered 95\% CL limit on $r$ is $r < \rlimitnoninety$.
The recovered value is  $\alens{} = \abbnoninety$ in this case; the slightly lower median value of \alens{} explains the equivalent limit on $r$ with less data.
Conversely, dropping the 150\,GHz autospectrum nearly doubles the uncertainty on \alens{} to \sigmaabbnoonefifty, and triples the limit on $r$ to $r < \rlimitnoonefifty$.

It is noticeable in Figure \ref{fig:SPTpol_bandpowers} that the 95\,GHz autospectrum appears consistently high, and in turn, as noted in \S\ref{sec:param_fitting}, the 95\,GHz Poisson power is the only clearly preferred foreground parameter. 
The default foreground model in this work independently floats the Poisson power in each frequency band, however a natural question is what spectral index is implied by the relative Poisson powers seen between the three bands. 
To answer this question, we tie together the Poisson terms, assuming that the source fluxes scale as a power law $\nu^{\alpha}$. 
Note that $\alpha$ is for the source fluxes, not the power. 
We set a uniform prior of $\alpha \in [-4,0]$. 
In this model, the upper limit on the Poisson power at 150\,GHz actually tightens by about 40\% (and zero is now excluded since the 95\,GHz result requires some power). 
The constraint on alpha is very weak with a 90\% CL range $\alpha \in [-3.2, -0.6]$; the posterior peaks at $\alpha \sim -2.65$. 
A spectral index in this range is on the low end of what has been observed for the temperature-selected sources in \citet{mocanu13}, but a large number of synchrotron sources in that work did have $\alpha < -0.6$. 
We also note that masking sources selected only at 150\,GHz (as in this work) would tend to drive this spectral index more negative. 
A alternative explanation is that the mean squared polarization fraction,  $\langle p^2 \rangle$, has increased going from 150 to 95\,GHz. 
The ratio of the median mean-squared polarization fraction at these two bands from \citet{gupta19} is 1.3; if one shifts these by one sigma in either direction, the ratio increases to 1.7.  
Such a shift in $\langle p^2 \rangle$ would have a similar effect on the 95 to 150\,GHz power as changing the spectral index by 0.6. 
It will be interesting to see if this frequency trend in the Poisson \bmode{} power holds up with future measurements, or if the excess power turns out to be the result of a systematic bias in the 95\,GHz bandpowers.

We also try fitting the data with doubled calibration and beam uncertainties.
As one would expect given the relative size of the bandpower error bars to the beam and calibration uncertainties, this has no impact on the recovered $r$ or \alens{} values.

Finally, we quantify the detection significance for cosmological BB power by looking at the probability for negative values of \alens{} in a MCMC run at high temperature in order to better sample the extreme tails of the posterior distribution. 
For the normal case, we find $P(\alens < 0) = \positiveBBwSVFGpte$, corresponding to a $ \positiveBBwSVFGsignif\,\sigma$ detection of positive \alens.
In the absence of sample variance, this probability becomes $P(\alens < 0) = \positiveBBwFGpte$, corresponding to a $\positiveBBwFGsignif\,\sigma$ detection of lensing \bmodes.
Lastly, to evaluate the detection significance of any \bmode{} power on the sky, we drop sample variance and set the foreground terms to zero.
We find $P(\alens < 0) = \positiveBBpte$, corresponding to a $\positiveBBsignif\,\sigma$ detection of any \bmode{} power.

\subsection{Constraints on primordial magnetic fields}
\label{subsec:pmf}

Measurements of the \bmode{} power spectrum also test models that predict primordial magnetic fields (PMFs) or cosmic birefringence \citep[e.g.,][]{kosowsky96}.
Both effects have the observational effect of rotating \emodes{} into \bmodes, thereby adding \bmode{} power compared to the standard \lcdm{} cosmology.
Since they are observationally indistinguishable in the bandpowers, we will only quote limits on the PMF power here.
These limits can be translated to apply to parity violating processes as well.

We follow the approach of \citet{sutton17}, drawing upon the vector and tensor PMF templates   for the CMB power spectra from that work.
We add these templates to the calculated CAMB spectra.
We assume the initial PMF anisotropy is Gaussian distributed with a nearly scale-invariant ($n_B=-2.9$) power law spectrum.
 Thus there are two free parameters: an overall power normalization $A_{PMF} \propto B_{1\,Mpc}^4$, and a timing parameter for when the PMF is generated $\beta = ln(a_\nu / a_{\rm PMF})$.
 Here, $B_{1\,Mpc}^4$ is the RMS strength of the PMF over 1\,Mpc scales, and $a_\nu$ and $a_{\rm PMF}$ are the scale factors at neutrino decoupling and PMF generation respectively.
 The chosen template is for $B_{1\,Mpc}=2.5$\,nG.
 The timing of PMF generation relative to neutrino decoupling impacts the magnitude of the tensor PMF modes.
 We follow earlier works \citep{planck15-30,zucca17,sutton17} and set a prior that $log_{10}(a_\nu / a_{\rm PMF}) \in [11.513, 41.447]$. 
We find the improved bandpower measurements in this work lead to a 95\% CL upper limit of $A_{PMF} < \apmf$.  
If we instead use the prior range of  $log_{10}(a_\nu / a_{\rm PMF}) \in [0, 16.937]$ from \citet{polarbear15}, the  95\% CL upper limit becomes $A_{PMF} < \apmflobeta$. 
As the bandpowers have scattered high (i.e.~$\alens{} = \abbrzero$ in \S\ref{sec:results}),  this upper limit from the BB bandpowers of this work alone is equivalent to the limit of $A_{PMF} < 0.36$ for $\alens=1$ that was found by \citet{sutton17} for the combination of \planck{}, POLARBEAR, BICEP2/Keck Array and the earlier 100\,\sqdeg{} SPTpol data releases. 


\section{Conclusions}
\label{sec:conclusion}

We present a measurement of the angular power spectrum of CMB \bmode{} polarization from the 500\,\sqdeg{} SPTpol survey.
Using three seasons of data, we report 21 bandpowers in seven multipole bins spanning $52 \le \ell \le 2302$ and three frequency combinations: 95 GHz $\times$ 95 GHz, 95 GHz $\times$ 150 GHz, and 150 GHz $\times$ 150 GHz.
These bandpowers represent the most precise direct measurement of \bmode{} power to date at small angular scales ($\ell > 320$), and range from angular scales where inflationary gravitational waves may be found to scales dominated by lensing \bmodes.

We have performed a strict set of null tests to probe the data for unknown systematic errors, and find no evidence for systematic contamination.
Astrophysical foreground \bmodes{} are a potential concern which we address by marginalizing over a Galactic template (important at low $\ell$) and independent Poisson power terms for each frequency band representing polarized extragalactic sources.
The Galactic template and prior is based on the measurements of Galactic polarized dust emission reported in \citet{bicep2keck18}. 
The data do not require these foreground terms, except in showing a 3\,$\sigma$ preference for Poisson power at 95\,GHz. 

Having found no evidence for systematic effects, we quantify the detection significance for astrophysical or cosmological \bmode{} power, and find the data rule out no \bmode{} power at \positiveBBsignif\,$\sigma$.
Marginalizing over astrophysical foregrounds but still neglecting sample variance, CMB \bmode{} power, consistent with the expectations for gravitational lensing, is detected at \positiveBBwFGsignif\,$\sigma$.
We check the data for consistency with the predicted lensing \bmodes{} in the \planck{} best-fit \lcdm{} model by fitting for an unknown rescaling \alens{} of the predicted lensing power.
We find $\alens{} = \abbrzero$, consistent with the expected value of unity in \lcdm.

With bandpowers extending down to $\ell=52$, this work is the first direct search for inflationary gravitational wave \bmodes{} with the South Pole Telescope.
The bandpowers presented here lead to a 95\% CL upper limit on the tensor-to-scalar ratio of $r < \rlimit$. 
This limit is close to what should be expected given the experimental sensitivity -- we calculate the expected $\sigma(r)$ for these measurement uncertainties is 0.22. 
This limit is largely set by the 150 $\times$ 150 GHz bandpowers due to the higher map noise level at 95\,GHz.
We can expect further improvements  from the new SPT-3G survey on the South Pole Telescope, a 1500 \sqdeg{} survey that began in 2018 and is expected to reach map depths a factor of several deeper than the data used here \citep{bender18}.

\acknowledgements{
The South Pole Telescope program is supported by the National Science Foundation through grant PLR-1248097.
Partial support is also provided by the NSF Physics Frontier Center grant PHY-0114422 to the Kavli Institute of Cosmological Physics at the University of Chicago, the Kavli Foundation, and the Gordon and Betty Moore Foundation through grant GBMF\#947 to the University of Chicago.  
This work is also supported by the U.S. Department of Energy. 
The Melbourne authors acknowledge support from an Australian Research Council Future Fellowship (FT150100074). 
J.W.H. is supported by the National Science Foundation under Award No. AST-1402161. 
W.L.K.W is supported in part by the Kavli Institute for Cosmological Physics at the University of Chicago through grant NSF PHY-1125897 and an endowment from the Kavli Foundation and its founder Fred Kavli. 
B.B. is supported by the Fermi Research Alliance LLC under contract no. De-AC02-07CH11359 with the U.S. Department of Energy.  
The Cardiff authors acknowledge support from the UK Science and Technologies Facilities Council (STFC).
The CU Boulder group acknowledges support from NSF AST-0956135.  
The McGill authors acknowledge funding from the Natural Sciences and Engineering Research Council of Canada, Canadian Institute for Advanced Research, and the Fonds de Recherche du Qu\'ebec -- Nature et technologies.
The UCLA authors acknowledge support from NSF AST-1716965 and CSSI-1835865.
Work at Argonne National Lab is supported by UChicago Argonne LLC,Operator  of  Argonne  National  Laboratory  (Argonne). Argonne, a U.S. Department of Energy Office of Science Laboratory,  is  operated  under  contract  no.   DE-AC02-06CH11357.  We also acknowledge support from the Argonne  Center  for  Nanoscale  Materials. 
This research used resources of the National Energy Research Scientific Computing Center (NERSC), a U.S. Department of Energy Office of Science User Facility operated under Contract No. DE-AC02-05CH11231.
The data analysis pipeline also uses the scientific python stack \citep{hunter07, jones01, vanDerWalt11} and the HDF5 file format \citep{hdf5}.
}

\bibliography{../../BIBTEX/spt}


\end{document}